# Multiscale computational approaches to magnetic behaviour in Cobalt Ferrite ($CoFe_2O_4$) nanostructures

Soham Chandra[a] and Soumyajit Sarkar[a,*]

[a]Department of Physics, Brainware University, Kolkata 700 125, INDIA

## ABSTRACT

Cobalt ferrite (***$CoFe_2O_4$***) is a prototypical ferrimagnetic spinel oxide whose exceptional magnetic anisotropy, magnetoelastic coupling, and thermal stability underpin applications in spintronics, magnetic hyperthermia, energy harvesting, and catalysis. This chapter presents a comprehensive computational framework that integrates electronic-structure calculations with atomistic spin modelling, statistical mechanics, and continuum micro-magnetics to predict magnetic functionality across length and time scales. Starting from density functional theory with Hubbard corrections (DFT+U), we derive exchange constants $J_{ij}$, magnetocrystalline anisotropy $K_1$, and magnetoelastic coefficients $B_1$, accounting for cation inversion, strain, and correlation effects. These parameters feed into generalized Heisenberg Hamiltonians, enabling Monte Carlo and Landau-Lifshitz-Gilbert simulations of finite-size effects, hysteresis, coercivity, and hyperthermia response in nanoparticles and thin films. Coarse-graining strategies bridge to micro-magnetic modelling, ensuring consistent parameter flow without empirical fitting. Computational case studies demonstrate size-dependent anisotropy enhancement, surface spin disorder, strain-tunable switching, and doping trends, revealing design principles inaccessible to experiment alone. Validation against benchmarks, e.g. Curie temperature, anisotropy constants, coercivity, magnetostriction, confirms predictive accuracy. Current challenges, e.g., U-parameter sensitivity, realistic surface chemistry, spin–lattice coupling, and large-scale integration are discussed alongside emerging directions including DFT+DMFT, coupled dynamics, and machine-learned potentials.

This multiscale hierarchy establishes cobalt ferrite as a quantitatively designable material, bridging microscopic electronic origins to device-relevant observables and providing the theoretical spine for magnetic functionality in this volume.

***KEYWORDS:***

Cobalt ferrite, Multiscale modelling, Magnetic anisotropy, DFT+U, Atomistic spin simulation, Magnetoelastic coupling

---

corresponding author: soumyajit.snb@gmail.com (SS)

# Contents

# Introduction

Cobalt ferrite ($CoFe_2O_4$) owes its technological prominence to its robust ferrimagnetism, exceptionally large magneto-crystalline anisotropy, and strong magneto-elastic coupling. These intrinsic characteristics directly govern the performance of the spintronic, magneto-elastic energy-harvesting, and biomedical systems explored throughout this volume. In spin valves and magnetic random - access memory (MRAM), anisotropy barriers and interfacial exchange control thermal stability and field- or current-driven switching. In strain-mediated energy harvesters, the modulation of magnetic anisotropy through applied stress forms the central tuning mechanism. For magnetic hyperthermia and targeted delivery, particle size, surface spin structure, and effective anisotropy together determine relaxation pathways and heating efficiency. At the microscopic level, the magnetism of $CoFe_2O_4$ arises from superexchange interactions between tetrahedral (A) and octahedral (B) sublattices, mediated by oxygen anions and governed by the Goodenough-Kanamori-Anderson rules (Anderson 1950, Goodenough 1958, Anderson 1987). The resulting ferromagnetic order, first formalised by Néel (Néel 1948), yields a net magnetisation whose magnitude and temperature dependence reflect the delicate balance among competing exchange pathways. In $CoFe_2O_4$, the dominant antiferromagnetic A-B coupling coexists with weaker intra-sublattice (B-B and A-A) interactions. The presence of $Co^{2+}$ ions on octahedral sites further introduces strong spin–orbit coupling, giving rise to a significant orbital contribution to the anisotropy (Coey 2010, Eskandari et al. 2017). Consequently, fundamental quantities such as the first anisotropy constant $K_1$, exchange stiffness, and saturation magnetization, $M_S$, originate directly from electronic-scale interactions rather than from purely phenomenological assumptions.

Experimental characterisation of these parameters in nanoscale systems, however, remains intrinsically difficult. Partial cation inversion-quantified by the inversion parameter, $\lambda$, modifies A-B exchange pathways and can appreciably shift both the Curie temperature and the anisotropy energy (Hou et al. 2010a, Hou et al. 2010b, Liu et al. 2019). Determining $\lambda$ in nanocrystalline samples is often indirect and model-dependent. Surface spin disorder adds further complexity: broken exchange bonds and reduced coordination at particle boundaries produce non-collinear spin configurations and a suppression of magnetisation that cannot be uniquely separated from bulk behaviour using magnetometry alone. In epitaxial thin films, strain- induced magnetoelastic coupling modifies the magneto-crystalline anisotropy, making it challenging to disentangle intrinsic electronic effects from extrinsic structural influences.

These experimental limitations have motivated the development of a predictive multiscale computational strategy capable of resolving magnetic interactions at their electronic origin and systematically propagating them across relevant length and time scales. Density functional theory, particularly within the GGA+U (or LSDA+U ) formalism, has become indispensable for treating the localised 3d states in transition-metal oxides and for extracting reliable sublattice magnetic moments and exchange constants in $CoFe_2O_4$ (Fritsch and Ederer 2012). When spin-orbit coupling is included, first-principles calculations can evaluate the magneto-crystalline anisotropy energy and its dependence on strain (Eskandari et al. 2017).

Such studies offer controlled insight into the influence of cation inversion, lattice distortion, and defect chemistry on intrinsic magnetic parameters. Electronic-structure methods alone, however, do not directly provide temperature-dependent observables or finite-size effects relevant to applications. Bridging this gap requires mapping the first principles-derived exchange interactions onto atomistic spin Hamiltonians, followed by statistical-mechanical treatment using Monte Carlo sampling or spin-dynamics simulations. Finite-size scaling techniques and Binder cumulant analysis (Binder 1981) allow reliable extraction of transition temperatures and critical behaviour in finite systems, while dynamical simulations capture switching and hysteresis phenomena beyond equilibrium thermodynamics. Only through this hierarchical approach can one connect Angstrom-scale exchange interactions to nanometre-scale coercivity and device-level thermal stability.

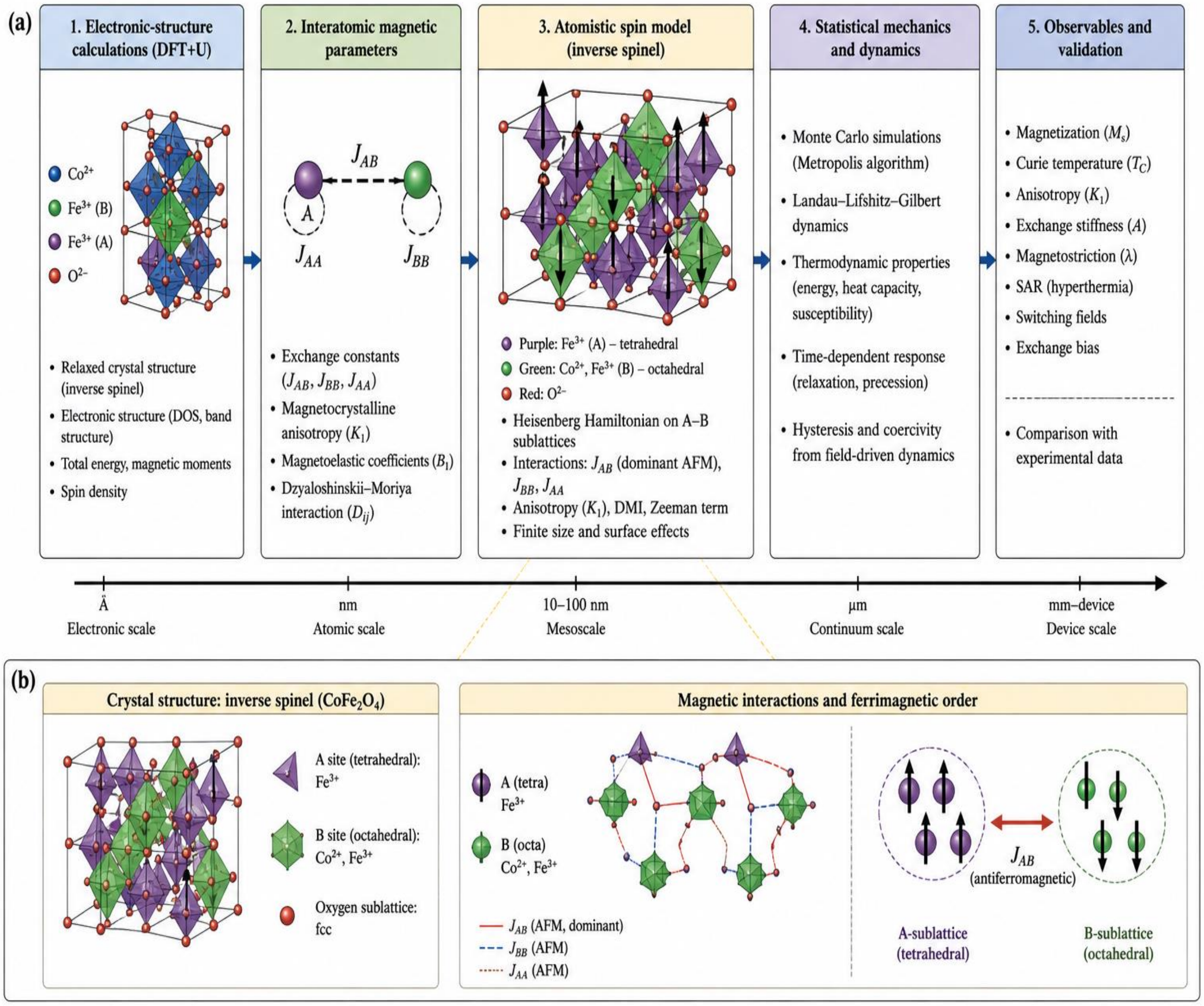

Figure 1: Multiscale computational framework for modeling magnetic properties of $CoFe_2O_4$ inverse spinel ferrites. (a) Workflow linking first-principles DFT+U calculations to atomistic spin modeling and statistical-mechanics simulations. Interatomic parameters (exchange interactions, anisotropy, magnetoelastic terms) are used to construct a spin Hamiltonian on the inverse spinel lattice, followed by Monte Carlo and spin-dynamics approaches to obtain thermodynamic and dynamic magnetic properties. (b) Crystal and magnetic structure of $CoFe_2O_4$, showing tetrahedral (A) and octahedral (B) sublattices. Dominant antiferromagnetic A–B exchange leads to ferrimagnetic ordering with antiparallel alignment of sublattice moments.

The present chapter does not revisit structural fundamentals or synthesis strategies– topics comprehensively covered in preceding chapters–but instead focuses on the computational integration of magnetic behaviour across scales. Its central aim is to establish a consistent parameter flow from electronic structure through atomistic spin models to mesoscale observables, with particular attention to transferability and proper scale separation. Although many studies have reported either isolated DFT calculations or phenomenological Monte Carlo simulations, rigorously parameterised multiscale treatments of $CoFe_2O_4$ nanostructures remain relatively scarce. The modelling gap addressed here therefore concerns hierarchical consistency: how to derive exchange and anisotropy parameters from first principles, incorporate realistic disorder and surface effects, and propagate these inputs into statistically robust simulations that quantitatively reproduce established experimental benchmarks such as Curie temperature, coercivity, and thermal stability (Franco and Silva 2010). Establishing such a predictive pipeline is essential for rationally connecting structural modifications—including inversion control, strain engineering, and doping—to quantitative magnetic functionality. A schematic overview of the multiscale computational framework, discussed in this work is shown in Fig. 1.

# 1. Ab Initio Description of Magnetic Interactions in $CoFe_2O_4$

## 1.1. Electronic Structure and Correlation Effects

A predictive description of magnetism in $CoFe_2O_4$ requires explicit treatment of its strongly correlated 3d electronic structure. The ferrimagnetic order arises from the partially filled 3d shells of $Co^{2+}$ and $Fe^{3+}$ ions distributed across octahedral (B) and tetrahedral (A) sites. In the inverse spinel configuration, $Fe^{3+}$ occupies both sublattices, while $Co^{2+}$ preferentially resides on octahedral B sites, giving rise to strong antiferromagnetic A–B superexchange. Because these 3d electrons are localised and experience significant on-site Coulomb repulsion, standard local or semi-local density functionals often underestimate correlation effects and may incorrectly predict metallic behaviour. The DFT+U method, introduced by Anisimov et al. (Anisimov et al. 1991) and later cast in a rotationally invariant form by Dudarev et al. (Dudarev et al. 1998), has therefore become the standard approach for spinel ferrites. The additional on-site Coulomb parameter U corrects self-interaction errors and restores the proper splitting between occupied and unoccupied d states. For $CoFe_2O_4$, both magnetic moments and exchange couplings display noticeable sensitivity to the chosen U values (Hou et al. 2010a, Hou et al. 2010b, Fritsch and Ederer 2012). Typical ranges employed in the literature are:

$$U_{Fe} = 3 - 5\ eV \text{ , and } U_{Co} = 4 - 6\ eV,$$

which yield magnetic moments in good agreement with experiment ( $\sim 5\ \mu B$ for $Fe^{3+}$ and a spin contribution of $\sim 3\ \mu B$ for $Co^{2+}$). Importantly, $Co^{2+}$ in octahedral coordination exhibits incomplete orbital quenching. When spin-orbit coupling (SOC) is included, an orbital moment of approximately $0.6 - 0.8\ \mu B$ appears on the Co sites (Fritsch and Ederer 2012, Rivas-Murias et al. 2023), contributing significantly to the magnetocrystalline anisotropy.

### 1.2. Extraction of Exchange Parameters

To build predictive atomistic spin models, the interatomic exchange constants $J_{ij}$ must be obtained from first principles. Within the magnetic force theorem, the Liechtenstein formalism (Liechtenstein et al. 1987) maps small rotations of magnetic moments onto a classical Heisenberg Hamiltonian:

$$H = -\sum_{i \neq j} J_{ij} \, \boldsymbol{S_i} \cdot \boldsymbol{S_j} \qquad \dots (1)$$

Alternatively, total energies of several collinear magnetic configurations can be computed and mapped onto effective exchange parameters. Both approaches consistently show that the antiferromagnetic A-B superexchange interaction dominates in $CoFe_2O_4$, in line with the Goodenough-Kanamori rules (Goodenough 1958). The intrasublattice B-B and A-A interactions are considerably weaker, yet they play an important role in determining spin-wave stiffness and finite-temperature behaviour.

First-principles studies indicate that the dominant A-B couplings exceed the intrasublattice interactions by nearly an order of magnitude, which readily accounts for the high Curie temperature ($T_C \sim 790\ K$) reported experimentally (Hou et al. 2010b, Fritsch and Ederer 2012). Nevertheless, these exchange parameters remain sensitive to lattice distortions, the degree of cation inversion, and the presence of defects.

### 1.3. Magnetocrystalline Anisotropy from First Principles

The magnetocrystalline anisotropy energy (MAE) is evaluated from relativistic DFT calculations by comparing total energies for different magnetisation directions. For the cubic case, the leading term takes the form:

$$E_{MAE} = K_1 \left(\alpha_1^2\alpha_2^2 + \alpha_2^2\alpha_3^2 + \alpha_3^2\alpha_1^2\right)$$

where $\alpha_i$ are the direction cosines of the magnetisation vector. DFT calculations consistently yield a large positive $K_1$, corresponding to $\langle 100 \rangle$ easy axes (Hou et al. 2010a, Fritsch and Ederer 2012). The microscopic origin of this anisotropy lies in the spin–orbit coupling that links the orbital moment of $Co^{2+}$ to the crystal field. Because anisotropy energies are small (typically of the order of $meV$ per formula unit), high numerical precision, and dense k-point sampling are essential to achieve reliable convergence.

### 1.4. Magnetoelastic Coupling and Strain Effects

Applied strain modifies the anisotropy through magnetoelastic coupling. First-principles calculations performed under controlled tetragonal distortion demonstrate that even modest biaxial strain can significantly alter the MAE (Odkhuu and Hong 2014). The magnetoelastic coefficient $B_1$ is readily extracted via the strain derivative:

$$B_1 = \frac{\partial E_{MAE}}{\partial \epsilon}$$

where $\epsilon$ denotes the strain. Strain also influences exchange interactions indirectly by changing Fe-O-Co bond angles and hopping amplitudes. Consequently, combined structural relaxation and spin–orbit-coupled calculations provide strain-dependent parameter sets that are indispensable for modelling thin-film spintronic devices.

### 1.5. Sensitivity to Cation Inversion and Defects

Cation inversion adds a further layer of complexity. The inversion parameter $\lambda$ alters exchange pathways and modifies the symmetry of the anisotropy. Supercell DFT calculations reveal that partial inversion affects both the sublattice magnetic moments and the hierarchy of $J_{ij}$ values (Hou et al. 2010b). By comparing total energies of different ordered configurations, it is also possible to estimate the thermodynamics of the inversion process. Oxygen vacancies, on the other hand, disrupt superexchange bridges and may introduce localised defect states. First-principles work shows that such vacancies can either enhance or suppress local magnetic moments, depending on the prevailing charge compensation mechanism.

The principal outcome of ab-initio modelling for $CoFe_2O_4$ is therefore a coherent microscopic parameter set:

$$\{J_{ij}, K_1, B_1, M_S, \text{and their dependence on strain and inversion}\}.$$

These quantities constitute the essential foundation for the atomistic spin Hamiltonians and finite-temperature simulations developed in the following sections. Without this electronic-structure underpinning, higher-level models would lack the predictive rigour required for quantitative materials design.

## 2. Construction of Atomistic Spin Hamiltonians

### 2.1. Generalized Heisenberg Hamiltonian

The first-principles parameters obtained in the preceding section supply the microscopic foundation for atomistic magnetic modelling. At length scales lying between the electronic structure and the continuum micromagnetic regime, the magnetic behaviour of $CoFe_2O_4$ is most naturally described by a generalised classical spin Hamiltonian defined on the discrete cation lattice. The starting point is the Heisenberg exchange term,

$$H_{ex} = -\sum_{i \neq j} J_{ij}\, \boldsymbol{S_i} \cdot \boldsymbol{S_j}$$

where the exchange constants $J_{ij}$ are derived from first-principles calculations (Hou et al. 2010b, Liechtenstein et al. 1987) and the classical spin vectors $\boldsymbol{S_i}$ are normalised such that $|\boldsymbol{S_i}| = 1$. In $CoFe_2O_4$, the dominant contribution comes from the antiferromagnetic A-B superexchange, while the weaker B-B and A-A interactions govern spin stiffness and finite-temperature fluctuations.

Magnetocrystalline anisotropy is incorporated through single-ion terms. For cubic symmetry the leading contribution reads

$$H_{ani} = -K_1 \sum_i (\alpha_{i1}^2\alpha_{i2}^2 + \alpha_{i2}^2\alpha_{i3}^2 + \alpha_{i3}^2\alpha_{i1}^2)$$

where, $\alpha_{ik}$ are the direction cosines of spin $\boldsymbol{S}_i$ with respect to the crystalline axes, and $K_1$ is obtained from spin-orbit-coupled DFT calculations (Hou et al. 2010a, Fritsch and Ederer 2012). In atomistic simulations of nanoparticles it is often convenient to replace the cubic term by an effective uniaxial anisotropy when the symmetry is lowered by shape or surface effects.

The Zeeman interaction with an external magnetic field $\boldsymbol{H}$ completes the minimal Hamiltonian:

$$H_Z = -\mu_0 \sum_i \mu_i \boldsymbol{S_i} \cdot \boldsymbol{H}$$

where $\mu_i$ denotes the sublattice magnetic moments taken from first-principles calculations.
The complete generalised Hamiltonian is therefore

$$H = H_{ex} + H_{ani} + H_Z$$

This discrete spin representation forms the computational framework for both Monte Carlo sampling and spin-dynamics simulations (Evans et al. 2014), allowing the calculation of key thermodynamic observables such as the temperature-dependent magnetisation $M(T)$, susceptibility $\chi(T)$, and coercivity.

## 2.2. Incorporating Disorder and Partial Inversion

Real $CoFe_2O_4$ samples rarely exhibit perfect inverse spinel order. Partial inversion is quantified by the inversion parameter, $\lambda$, which gives the fraction of $Co^{2+}$ ions occupying tetrahedral A sites. At the atomistic level this requires stochastic assignment of site occupancies. A practical approach is to construct large supercells in which A- and B-site occupations are assigned according to $\lambda$, either randomly or according to energetically weighted probabilities derived from DFT total-energy differences (Hou et al. 2010a, b). The exchange constants $J_{ij}$ then become locally environment-dependent, reflecting the actual chemical surroundings.

Disorder introduces spatial fluctuations in exchange strengths, effectively replacing a single $J_{ij}$ value by a distribution, $P(J_{ij})$. Such randomness tends to broaden magnetic phase transitions and lower the effective Curie temperature, in agreement with finite-size scaling theory (Binder 1981). In Monte Carlo simulations, proper disorder averaging over multiple independent configurations is therefore essential for statistically reliable results. Vacancies and non-stoichiometric defects can be included by selectively removing exchange bonds or modifying local anisotropy constants. Oxygen vacancies, for instance, break Fe-O-Co superexchange paths and reduce the local $J_{ij}$ values.

## 2.3. Surface and Finite-Size Effects

In nanoparticles and thin films, surface effects become particularly prominent. Broken exchange bonds at the surface reduce the coordination number and weaken the effective exchange field. In atomistic models, this is implemented simply by truncating the exchange interactions at the particle boundary. Surface spins also experience modified anisotropy owing to the reduced local symmetry. A widely used approach is to introduce an additional surface anisotropy term:

$$H_{surf} = -K_s \sum_{i \in surface} (\boldsymbol{S}_i \cdot \boldsymbol{n}_i)^2$$

where $K_S$ is the surface anisotropy constant and $\boldsymbol{n}_i$ is the local surface normal (Omelyanchik et al. 2020). This term can compete with the bulk cubic anisotropy and often induces spin canting. The competition between strong A-B exchange in the core and weakened interactions at the surface naturally leads to core-shell magnetic structures: an ordered ferrimagnetic core coexisting with a disordered or canted surface shell. Atomistic simulations successfully reproduce the experimentally observed reduction in saturation magnetisation and the enhancement of coercivity in nanoscale $CoFe_2O_4$ particles (Evans et al. 2014). Finite size also influences thermal stability. In small particles, magnetisation reversal may proceed via coherent rotation or non-uniform modes, depending on the particle diameter relative to the exchange length. Atomistic models capture this crossover directly, without the need to impose continuum approximations.

## 2.4. Parameter Transferability and Renormalization

Although the exchange constants $J_{ij}$ and anisotropy $K_1$ are extracted from bulk DFT calculations, their direct application to nanoscale systems must be treated with care. Surface relaxation, strain gradients, and spatially varying inversion can modify the local environment and renormalise the magnetic parameters. Temperature further renormalises both effective anisotropy and exchange stiffness. Within classical spin models, these effects emerge naturally from thermal fluctuations. Nevertheless, it is often useful to introduce an explicit temperature-dependent anisotropy of the form:

$$K_1(T) \propto [M(T)]^n$$

Where, $n \approx 3$ for single-ion anisotropy in localised spin systems (Callen 1963).

True parameter transferability therefore demands consistency checks between bulk-derived values and nanoscale observables. When discrepancies appear, site-dependent corrections or coarse-graining procedures that preserve thermodynamic behaviour can be employed. Bridging the atomistic and micromagnetic scales involves extracting continuum parameters such as the exchange stiffness $A$ and an effective anisotropy constant $K_{eff}$ directly from the atomistic simulations. This mapping guarantees hierarchical coherence throughout the multiscale framework.

## 2.5. Limitations of Classical Spin Approximations

(a) Despite their considerable utility, classical spin Hamiltonians have intrinsic limitations. They neglect quantum fluctuations, which can be relevant at low temperatures. Moreover, mapping itinerant-electron effects onto localised spin models assumes well-defined magnetic moments on every site - an approximation that is reasonable for $CoFe_2O_4$ but not universally valid.

(b) A second limitation concerns longitudinal spin fluctuations. Classical models usually fix the spin magnitude, thereby omitting amplitude variations that become important near $T_C$. Extensions based on Landau–Lifshitz–Bloch dynamics offer a partial remedy, albeit at higher computational cost (Garanin 1997). Finally, the exchange parameters themselves inherit the uncertainty of the chosen U values in DFT+U calculations. This sensitivity propagates into the atomistic regime and underscores the need for careful parameter sensitivity analysis.

The atomistic Hamiltonian constructed in this section forms the crucial intermediate bridge between the electronic-structure description and macroscopic magnetic observables. It incorporates first-principles-derived exchange and anisotropy parameters, explicitly treats disorder and finite-size effects, and serves as the computational engine for all temperature-dependent modelling that follows. The reliability of every higher-level simulation presented in this chapter ultimately rests on the rigour and internal consistency of this construction.

# 3. Statistical Mechanics of Nanoscale Ferrimagnetism

## 3.1. Monte Carlo Framework and Equilibrium Thermodynamics

Having established the generalized atomistic spin Hamiltonian,

$$H = -\sum_{i \neq j} J_{ij} \boldsymbol{S_i} \cdot \boldsymbol{S_j} + H_{ani} + H_Z$$

Finite-temperature properties of ferrimagnetic spinels such as $CoFe_2O_4$ are obtained through statistical sampling of spin configurations. The canonical ensemble distribution is:

$$P(\{S_i\}) \propto \exp\left(-\frac{H}{k_B T}\right)$$

With $k_B$ the Boltzmann constant and $T$ temperature. Metropolis Monte Carlo methods (Metropolis et al. 1953) remain the standard numerical tool for classical spin systems. A trial rotation of a randomly selected spin is accepted with probability: $P_{acc} = \min\left\{1, \exp\left(-\frac{\Delta E}{k_B T}\right)\right\}$ ensuring importance sampling of thermodynamically relevant configurations.

For $CoFe_2O_4$, explicit treatment of both tetrahedral (A) and octahedral (B) sublattices is essential to preserve ferrimagnetic A-B superexchange originally described by Néel (Néel 1948). Separate monitoring of sublattice magnetizations:

$$M_A = \frac{1}{N_A} \langle \textstyle\sum_{i \in A} S_i^z \rangle \quad \text{and, } M_B = \frac{1}{N_B} \langle \textstyle\sum_{i \in B} S_i^z \rangle$$

reveals temperature-dependent compensation behavior in doped or inversion-modified systems. Magnetic susceptibility follows from fluctuations:

$$\chi = \frac{N}{k_B T} \left( \langle M^2 \rangle - \langle M \rangle^2 \right)$$

while specific heat derives from energy variance,

$$C = \frac{N}{k_B T^2} \left( \langle E^2 \rangle - \langle E \rangle^2 \right)$$

Near the transition region, critical slowing down necessitates extensive equilibration (typically $10^4 - 10^5$ MCS per spin). Surface disorder in nanoparticles further increases equilibration time due to metastable states separated by anisotropy barriers.

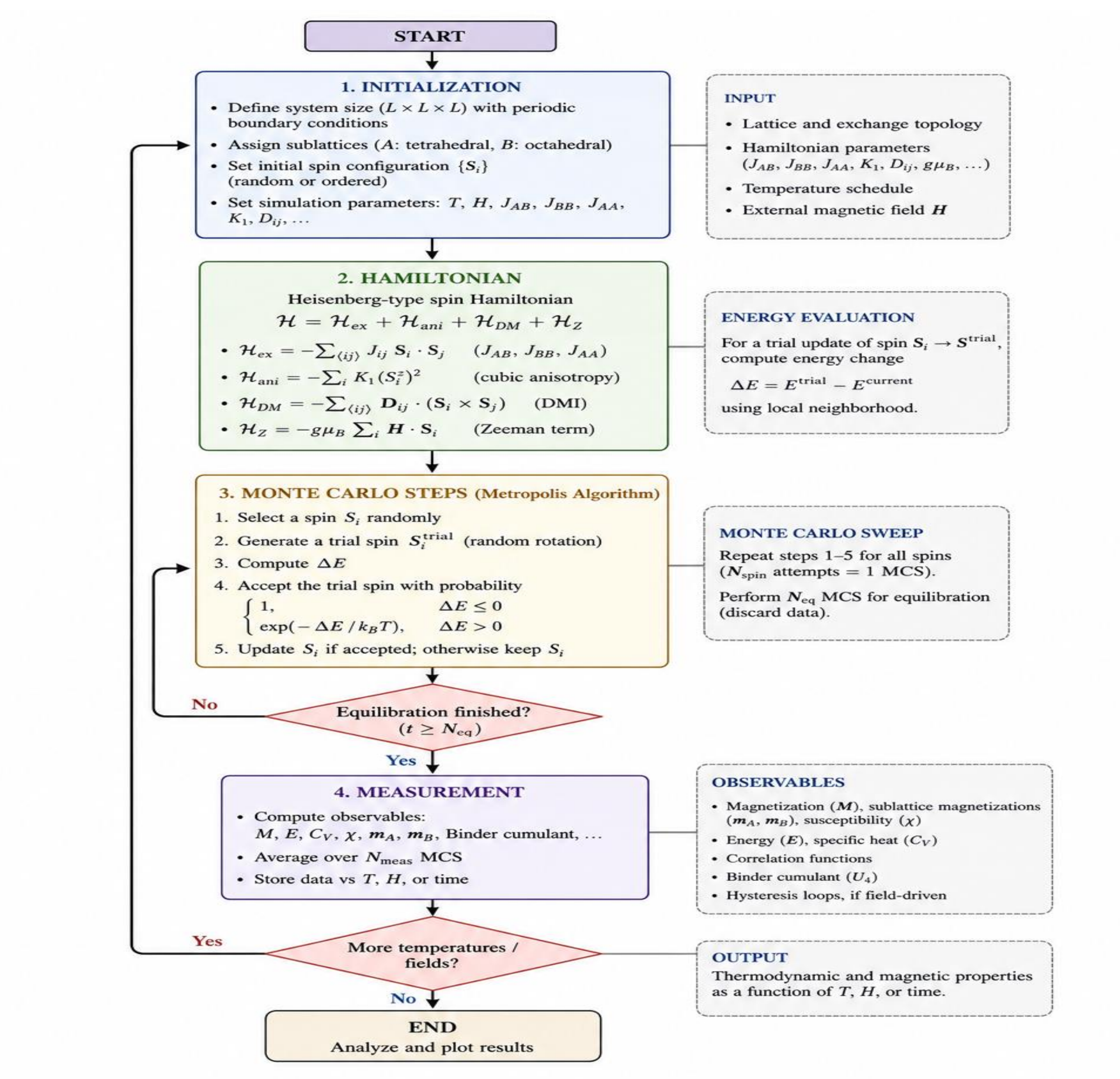

Figure 2: Monte Carlo simulation workflow for atomistic spin models of $CoFe_2O_4$ inverse spinel ferrites. The procedure includes initialization of the lattice and model parameters, construction of the spin Hamiltonian, Metropolis-based Monte Carlo updates with energy evaluation, and measurement of thermodynamic and magnetic observables after equilibration.

### 3.2. Finite-Size Scaling and Determination of the Curie Temperature

Direct identification of $T_C$ from magnetization collapse yields size-dependent values. Instead, Binder cumulant analysis (Binder 1981) provides a reliable estimate:

$$U_L = 1 - \frac{\langle M^4 \rangle}{3\langle M^2 \rangle^2}$$

Curves of $U_L(T)$ for different lattice sizes intersect at the thermodynamic Curie temperature. Bulk-like $CoFe_2O_4$ simulations employing first-principles-derived exchange parameters reproduce experimental Curie temperatures $(\sim 790\,K)$ when appropriate correlation corrections are used (Szotek et al. 2006). Deviations highlight sensitivity of $J_{ij}$ to electronic structure methodology. Finite-size scaling (Binder 1981) predicts: $T_C(L) = T_C(\infty) - aL^{-\frac{1}{\nu}}$, and for nanoparticles of diameter $D$: $T_C(D) = T_C(\infty)\left(1 - \frac{a}{D^{1/\nu}}\right)$. Reduced coordination and enhanced surface spin disorder suppress long-range ordering in nanoscale $CoFe_2O_4$, explaining experimentally observed $T_C$ reduction.

### 3.3. Hysteresis, Coercivity, and Energy Barrier Landscape

Quasi-static hysteresis loops are obtained by sweeping an external magnetic field at fixed temperature. At each field value, the system is equilibrated and magnetization recorded, generating $M(H)$ curves. The coercive field $H_C$ and remanent magnetization $M_r$ depend strongly on magneto-crystalline anisotropy $K_1$, particle size, and disorder. The large anisotropy of $CoFe_2O_4$ enhances coercivity relative to soft ferrites. Surface anisotropy further increases effective energy barriers in nanoparticles.

Switching behavior may be interpreted through the energy barrier, $\Delta E \approx K_{eff}V$, competing with thermal energy $k_B T$. Monte Carlo sampling reveals metastable states and nonuniform reversal pathways, particularly when exchange randomness arises from partial cation inversion. Because conventional Monte Carlo lacks intrinsic time evolution, field sweep rate does not correspond to physical time. Quantitative switching dynamics therefore require dynamical simulations.

### 3.4. Spin Dynamics and the Landau–Lifshitz–Gilbert Equation

Time-dependent magnetization processes are modeled via atomistic spin dynamics, solving the Landau-Lifshitz-Gilbert (LLG) equation (Gilbert 2004, Bertotti 2004):

$$\frac{d\boldsymbol{S}_i}{dt} = -\gamma \boldsymbol{S}_i \times \boldsymbol{H}_i^{eff} + \alpha \boldsymbol{S}_i \times \frac{d\boldsymbol{S}_i}{dt}$$

where γ is the gyromagnetic ratio and α the Gilbert damping parameter. The effective field is

$$\boldsymbol{H}_i^{eff} = -\frac{1}{\mu_i}\frac{\partial H}{\partial \boldsymbol{S}_i}$$

Thermal fluctuations are incorporated through a stochastic field satisfying the fluctuation-dissipation theorem (Brown 1963a). This stochastic LLG equation enables simulation of thermally activated reversal and dynamic hysteresis. In $CoFe_2O_4$ nanoparticles, very small particles exhibit near-coherent rotation, while larger particles display non-uniform reversal involving surface nucleation and domain-wall-like structures. The damping parameter, $\alpha$, controls relaxation time and switching speed-parameters central to MRAM and spin-transfer torque applications.

### 3.5. Magnetic Relaxation and Superparamagnetism

When particle size becomes sufficiently small, thermal fluctuations overcome anisotropy barriers, leading to superparamagnetism. The relaxation time follows the Néel-Arrhenius law:

$$\tau = \tau_0 \exp\left(\frac{K_{eff}V}{k_B T}\right)$$

originally derived by Néel for fine magnetic grains (Néel 1949) and later refined in the classical single-domain treatment (Brown 1963a). The blocking temperature $T_B$ is defined by the condition $\tau(T_B) \approx \tau_m$, where $\tau_m$ is the experimental measurement time. Enhanced surface anisotropy in $CoFe_2O_4$ can significantly increase $K_{eff}$, stabilizing magnetization in nanoscale particles. Atomistic free-energy calculations quantify reversal barriers and clarify discrepancies between experimental techniques operating on different time scales.

### 3.6. Dynamic Susceptibility and Hyperthermia Response

Under alternating magnetic fields, energy dissipation arises from both hysteresis losses and relaxation processes. Spin-dynamics simulations can compute the dynamic susceptibility $\chi(\omega)$ as well as the absorbed power per unit volume. The hysteresis-loss contribution to the power density is given by:

$$P = \mu_0 f \int M(H) dH \text{ ,}$$

where $f$ is the frequency of the applied field and the integral represents the area enclosed by the dynamic hysteresis loop (Carrey et al. 2011). (Equivalently, when working with angular frequency $\omega = 2\pi f$, the expression reads,

$$P = \frac{\mu_0 \omega}{2\pi} \int M(H) dH \text{ .}$$

.Monte Carlo and spin-dynamics simulations transform atomistic exchange and anisotropy parameters into temperature dependent magnetization curves, coercivity trends, blocking temperatures, and dynamic susceptibilities. These outputs support MRAM stability criteria, spin-valve switching thresholds, and energy-harvesting performance metrics. When exchange constants and anisotropy parameters originate from first-principles calculations without empirical adjustment, the multiscale hierarchy becomes internally consistent. Structural modifications-strain, inversion tuning, or defect engineering-can then be translated quantitatively into macroscopic magnetic functionality in $CoFe_2O_4$ and related ferrimagnetic nanostructures.

# 4. Multiscale Coupling: From Electronic Structure to Device-Relevant Observables

The predictive power of computational magnetism does not arise from any single method, but from the coherent coupling of approaches across electronic, atomistic, mesoscopic, and device scales. For technologically relevant ferrimagnetic oxides such as $CoFe_2O_4$, magnetic functionality emerges through a natural cascade of interactions: correlated 3d electrons determine the exchange integrals; these exchange parameters, together with anisotropy, govern thermodynamic ordering; collective spin textures control switching behaviour; and the resulting switching characteristics ultimately dictate device performance. This section develops the multiscale hierarchy that systematically connects first-principles electronic structure to experimentally measurable and device-relevant observables. By emphasising consistent parameter flow and proper scale separation, the present treatment distinguishes itself from purely structural discussions or application-focused chapters in this volume.

## 4.1. Hierarchy of Length and Time Scales

Magnetic phenomena in $CoFe_2O_4$ span wide spatial and temporal domains:

- Length scales: Å (electronic orbitals) → nm (atomic clusters and nanoparticles) → 10-100 nm (domains and patterned elements).
- Time scales: fs (electronic hopping and spin–orbit interactions) → ps-ns (precession and damping) → μs and beyond (thermal relaxation and device switching).

At the Angstrom scale, DFT calculations with Hubbard corrections (Anisimov et al. 1991, Dudarev et al. 1998) determine band structure, exchange pathways, and magnetocrystalline anisotropy. These microscopic quantities define the Hamiltonian that governs spin behaviour at the nanometre scale. At the nanometre scale, atomistic spin simulations describe thermodynamic ordering, finite-size effects, and reversal mechanisms in $CoFe_2O_4$ nanoparticles and thin films. Moving toward tens of nanometres, micromagnetic modelling captures domain-wall formation, non-uniform magnetisation textures, and magnetostatic interactions. Finally, at device scales (∼100 nm and above), transport, magnetoelastic, and thermal effects must be incorporated self-consistently. A coherent multiscale strategy ensures that information flows smoothly across these regimes without empirical disconnects.

## 4.2. Parameter Flow Across Scales

### 4.2.1. DFT → Atomistic Spin Models

Electronic-structure calculations provide the fundamental exchange constants, $J_{ij}$, single-ion anisotropy parameters $K$, and magnetoelastic coefficients. In correlated oxides such as $CoFe_2O_4$, the LSDA+U approach (Anisimov et al. 1991, Dudarev et al. 1998) is essential for capturing the correct insulating behaviour and the proper distribution of magnetic moments. First-principles studies demonstrate that cation inversion and strain significantly modify exchange pathways (Hou et al. 2010b, Fritsch and Ederer 2012). These $J_{ij}$ values populate the atomistic Heisenberg

Hamiltonian used in Monte Carlo and spin-dynamics simulations. The Curie temperature, sublattice magnetisation, and susceptibility then emerge naturally from statistical sampling.

Complementary electronic-structure work using the self-interaction corrected local spin density approximation (SIC-LSDA) has examined both inverse and normal spinel structures (Szotek et al. 2006). Both configurations are insulating, although the normal spinel exhibits smaller band gaps. The calculated spin magnetic moments and the exchange splitting of the conduction bands increase when moving from the inverse to the normal spinel structure. More recent investigations employing PBE-GGA+U have explored different sub-magnetic orderings in $CoFe_2O_4$, determining band structures, densities of states, and exchange coupling energies via the magnetic force theorem (Duru 2022). These calculations reveal stronger Co-O hybridisation compared with Fe-O, varying band gaps across magnetic states, and magnetocrystalline anisotropy values around $0.092\ eV$. The resulting DFT-derived parameters (exchange energies, anisotropy constants, and magnetic moments) serve as direct inputs for Monte Carlo simulations on a classical Heisenberg model, yielding Curie temperatures up to $\sim 725\ K$ for the ferromagnetic sub-ordered state, with other configurations transitioning at lower temperatures (below 500 K and 445 K), consistent with variations observed in experimental synthesis conditions.

### 4.2.2. Atomistic → Mesoscale (Micromagnetics)

From experimental studies, we know spin reorientation in $CoFe_2O_4$ thin single crystalline films epitaxially grown on (100) MgO substrate, occurs at a critical thickness of 300 nm upon varying the film thickness, driven by structural transition from tetragonal to cubic symmetry. For low film thicknesses, the in-plane tensile stress induces tetragonal distortion of the crystal lattice, generating a perpendicular anisotropy of sufficient magnitude to overcome the demagnetizing (shape) anisotropy and stabilize an out-of-plane magnetization easy axis. In thicker films, progressive lattice relaxation toward the cubic bulk structure enables the shape anisotropy to prevail, reorienting the easy axis back into the film-plane (Lisfi et al. 2007)

Atomistic simulations yield effective anisotropy constants, temperature-dependent magnetization $M(T)$, and exchange stiffness $A$. These are coarse-grained to continuum parameters entering the Landau-Lifshitz-Gilbert framework at the micromagnetic scale. For example, exchange stiffness relates to atomistic exchange via:

$$A = \frac{1}{2}\sum_{j} \frac{J_{ij} r_{ij}^2}{V_{cell}}$$

Where, $r_{ij}$ is interatomic separation and $V_{cell}$ the volume per spin (adjusted for multi-sublattice ferrimagnets by averaging dominant contributions) (Atxitia et al. 2010). Atomistic spin-wave dispersion fitting $(\omega(k) = Dk^2)$ provides an alternative: $A = \frac{DM_s}{2\gamma}$ (Waring et al. 2023).

Effective anisotropy, $K_{eff}$, is renormalized by thermal fluctuations and surface effects. The Callen-Callen law predicts: $K_{eff}(T) \propto [M(T)]^3$, for single-ion anisotropy (Callen 1963, Moreno et al. 2025). In nanoparticles, surface contributions yield: $K_{eff} = K_{bulk} + \frac{6K_s}{D}$ (Pujada et al. 2003,

Mumtaz et al. 2007). Thus, micromagnetic simulations of patterned $CoFe_2O_4$ elements inherit physically grounded parameters rather than phenomenological fits.

### 4.3. Coarse-Graining and Renormalization Strategies

Direct simulation of device-scale systems with atomistic resolution is computationally prohibitive. Coarse-graining strategies reduce degrees of freedom while preserving essential physics. One approach averages clusters of spins into macrospins, renormalizing exchange and anisotropy to preserve long-wavelength excitations (Evans et al. 2014). Renormalization group concepts justify scale-dependent effective parameters, particularly near criticality (Grinstein and Koch 2003).

Temperature renormalization is crucial: effective anisotropy $K_{eff}(T)$ decreases with temperature due to spin fluctuations, often following power-law behavior. Ignoring such renormalization can lead to overestimation of coercivity in device simulations. For $CoFe_2O_4$, strong magnetocrystalline anisotropy simplifies coarse-graining by stabilizing near-uniform local alignment, yet surface and inversion disorder introduce spatial inhomogeneities requiring careful averaging.

### 4.4. Magnetostatic Interactions and Micromagnetic Dynamics

Continuum micromagnetics incorporates long-range dipolar interactions explicitly. The magnetostatic energy is:

$$E_{demag} = -\frac{1}{2}\mu_0 \int \boldsymbol{M} \cdot \boldsymbol{H_d}\, dv$$

where $H_d$ is the demagnetizing field. Competition between exchange, anisotropy, and magnetostatics determines domain structure. The characteristic exchange length,

$$l_{ex} = \sqrt{\frac{2A}{\mu_0 M_s^2}}$$

sets the natural length scale for domain-wall width and vortex formation (Brown 1963b). For $CoFe_2O_4$, large $K_1$ and moderate $M_s$ produce relatively narrow domain walls, consistent with its high coercivity. Continuum dynamics follow the Landau-Lifshitz-Gilbert equation:

$$\frac{\partial \boldsymbol{M}}{\partial t} = -\gamma \boldsymbol{M} \times \boldsymbol{H_{eff}} + \frac{\alpha}{M_s} \boldsymbol{M} \times \frac{\partial \boldsymbol{M}}{\partial t}$$

With, $\boldsymbol{H_{eff}} = -\frac{1}{\mu_0}\frac{\delta E}{\delta \boldsymbol{M}}$. Solvers such as OOMMF (Donahue and Porter 1999) or MuMax3 (Vansteenkiste et al. 2014) enable simulation of patterned films, domain-wall motion, and spin-torque effects using parameters from atomistic modeling.

### 4.5. Validation of the Multiscale Framework and Integration into Application-Oriented Modeling

Rigorous validation of the proposed multiscale approach requires quantitative comparison against key experimental observables across length and time scales.

- Curie temperature ($T_C$): Monte Carlo simulations, parameterized with first-principles exchange interactions $J_{ij}$, yield a bulk $T_C \approx 790 - 820\ K$ for $CoFe_2O_4$, in good agreement with experimental values for bulk material (~793–825 K) and recent atomistic predictions.
- Magnetocrystalline anisotropy constant ($K_1$): Density functional theory calculations predict strain-dependent trends in $K_1$, consistent with experimental torque magnetometry data on epitaxial thin films and bulk systems (Fritsch and Ederer 2012).
- Coercivity($H_C$): Finite-temperature micromagnetic simulations incorporating renormalized, temperature-dependent anisotropy and realistic nanoparticle geometries reproduce experimental $H_C$ trends, including size-dependent maxima and thermal decay observed in $CoFe_2O_4$ nanoparticles (Lamouri et al. 2020).
- Magnetostriction and magnetoelastic response: First-principles magnetoelastic coefficients enable accurate computation of strain–magnetization coupling, aligning with measured piezomagnetic coefficients and strain-induced anisotropy modulation (Fritsch and Ederer 2012).

This multi-faceted agreement across thermal, structural, and magnetic benchmarks establishes the predictive reliability of the framework. The validated parameters and effective models are subsequently integrated into application-oriented simulations:

- Spin transport in magnetic tunnel junctions and MRAM: DFT-derived exchange splitting and spin polarization inform tunneling magnetoresistance calculations, while micromagnetic textures (domain walls, vortices) serve as input configurations for micromagnetic-transport solvers.
- Magnetoelastic and strain-mediated devices: Combining first-principles magnetoelastic tensors, atomistic temperature renormalization of anisotropy, and continuum elasticity modeling enables prediction of strain-tunable magnetic switching and energy-harvesting performance in low-power, voltage-controlled spintronic and multiferroic heterostructures.

This integrated workflow bridges ab initio accuracy with device-scale functionality, facilitating design optimization for emerging magnetic technologies.

***Distinctive Role of Multiscale Coupling in Predictive Modeling***

In contrast to structural studies centered on crystal chemistry or device-oriented chapters focused on architectures, multiscale coupling provides a causal continuum across length and energy scales: electronic correlations dictate microscopic exchange interactions; these interactions govern thermodynamic ordering and phase stability; macroscopic order in turn controls magnetization reversal and switching dynamics; and switching ultimately dictates device-level performance metrics. For $CoFe_2O_4$, this hierarchy is particularly transparent owing to the material's high tunability through cation inversion degree, epitaxial strain, and surface/interface engineering. The consistent upward propagation of electronic-structure-derived parameters (e.g., exchange constants $J_{ij}$, anisotropy $K_1$, and magnetoelastic coefficients) through atomistic Monte Carlo and mesoscale micromagnetic simulations enables quantitatively predictive modeling. This approach transforms cobalt ferrite from an empirically characterized material into a rationally designable functional oxide for applications in spintronics, magnetoelastic devices, and multiferroic heterostructures.

# 5. Computational Case Studies

## 5.1. Nanoparticles: Biomedical and Catalytic Context

### 5.1.1. Size-Dependent Magnetic Anisotropy

When the particle diameter D approaches the magnetic exchange length, the magnetic response of $CoFe_2O_4$ deviates significantly from bulk behavior. Atomistic simulations parameterized using first-principles exchange constants $J_{ij}$ and anisotropy constants $K$ show systematic enhancement of the effective anisotropy with decreasing size.

Within a core-surface decomposition framework,

$$K_{eff}(D) = K_{bulk} + \frac{6K_S}{D} \quad \ldots (2)$$

Where $K_S$ denotes the surface anisotropy density. First-principles calculations indicate that $K_S$ depends sensitively on surface termination and local inversion degree (Fritsch and Ederer 2012). Monte Carlo simulations incorporating these parameters reproduce experimentally observed coercivity enhancement in the 8–20 nm size regime (Evans et al. 2014).

### 5.1.2. Surface Spin Disorder and Magnetization Reduction

Broken A–B superexchange bonds at the surface reduce local exchange fields and induce spin canting. Atomistic modeling reveals formation of a core–shell spin structure consisting of an ordered ferrimagnetic core and a disordered surface shell. The reduction in saturation magnetization can be approximated as (Dutta et al. 2009):

$$M_S(D) = 92\left(1 - \frac{2\delta}{D}\right)^3 \quad \ldots (3)$$

where $\delta$ represents an effective shell thickness, though actual shell thicknesses vary (1.9 - 6.8 Å) depending on material and synthesis method. Monte Carlo simulations within the Ising model and free boundaries conditions on cubic geometry with different nanoparticle sizes show significant impact of particle size on magnetic properties of Cobalt-ferrite nanoparticles (Lamouri et al. 2020), and reproduce the experimentally observed reduction in $M_S$ for particle sizes below $\sim$ 10 nm.

### 5.1.3. Implications for Magnetic Hyperthermia

For hyperthermia applications, heating efficiency depends on hysteresis losses or N´eel relaxation dynamics. Spin-dynamics simulations indicate that optimal heating performance occurs when

$$K_{eff}V \approx 25\, k_B T \quad \ldots (4)$$

ensuring appropriate relaxation times within biomedical frequency windows (Carrey et al. 2011). Excessively large anisotropy suppresses relaxation and reduces specific absorption rate (SAR), highlighting the importance of quantitative optimization via tunable surface parameters.

### 5.1.4. Magnetic Structure and Catalytic Activity

First-principles studies demonstrate that surface spin polarization modifies adsorption energetics via spin-dependent charge redistribution. Surface cation inversion alters Fe-O covalency and oxygen vacancy formation energies, linking magnetic order to catalytic performance in oxygen evolution/reduction reactions.

## 5.2. Thin Films: Spintronics Context

### 5.2.1. Epitaxial Strain and Anisotropy Control

In epitaxial thin films, biaxial strain ε acts as a tunable thermodynamic variable. Spin-orbit coupled density functional theory calculations show that tetragonal distortion significantly modifies magnetocrystalline anisotropy energy (MAE) (Fritsch and Ederer 2012). For small strain,

$$K(\varepsilon) = K_0 + B_1(\varepsilon) \qquad \dots (5)$$

where $B_1$ is the magnetoelastic coefficient extracted from total-energy calculations (Hou et al. 2010a, b). $CoFe_2O_4$ exhibits comparatively large $B_1$, explaining strong strain tunability observed experimentally.

### 5.2.2. Strain-Inversion Coupling

Partial inversion modifies $Co^{2+}$ orbital occupancy, thereby altering the strain dependence of MAE. Micromagnetic simulations incorporating DFT-derived parameters reproduce experimentally measured switching fields in CFO-based heterostructures, demonstrating how combined strain and inversion engineering enables anisotropy control for spintronic devices.

## 5.3. Doped Systems

### 5.3.1. Exchange Modification Trends

Substitutional doping modifies superexchange pathways. $Zn^{2+}$ preferentially occupies $A$ sites, reducing $J_{AB}$ and lowering Curie temperature. $Mn^{2+}$ substitution enhances magnetic moment but weakens anisotropy due to reduced spin–orbit coupling. Atomistic simulations incorporating modified exchange matrices reproduce experimentally observed $T_C$ trends (Shaikh et al. 2015).

### 5.3.2. Rare-Earth-Induced Anisotropy Enhancement

Rare-earth substitution (e.g., $Dy^{3+}$) introduces strong localized single-ion anisotropy contributions due to unquenched orbital moments. Even modest substitution levels (5-10%) can increase effective anisotropy by 20-40% in modeled systems (Li et al. 2022).

## 5.4. Predictive Design Guidelines

The case studies collectively illustrate a multiscale modeling pipeline:

$$\text{DFT} \rightarrow \{J_{ij}, K, B_1\} \rightarrow \text{Spin Hamiltonian} \rightarrow \text{Finite-T simulations} \rightarrow \text{Device observables.}$$

Simulations uniquely enable independent control of inversion degree, strain, dopant concentration, and surface exchange parameters—variables that experiments cannot isolate simultaneously. Across nanoparticles, thin films, and doped systems, magnetic anisotropy emerges as the primary determinant of functional performance (coercivity for hyperthermia, strain-tunability for spintronics, enhanced barriers for stability). However, most computational studies treat inversion, strain, and defect effects independently. A predictive framework must incorporate their nonlinear coupling within a unified statistical-mechanical description. Such integration reveals design rules-e.g., optimal inversion/strain combinations for maximized SAR or switching speed-that guide synthesis and device optimization beyond empirical trial-and-error.

## 6. Current Challenges and Emerging Computational Directions

### 6.1. Accurate Treatment of Strong Correlation

Although density functional theory augmented with on-site Hubbard corrections (DFT+U) has been widely employed for $CoFe_2O_4$, the quantitative reliability of predicted exchange constants and magnetocrystalline anisotropy energies remains sensitive to the chosen U parameters (Anisimov et al. 1991, Dudarev et al. 1998). The partially filled 3d shells of $Co^{2+}$ and $Fe^{3+}$ exhibit moderate to strong electronic correlation, and static mean-field corrections may not fully capture multiplet structure and dynamical fluctuations.

First-principles studies have shown that magnetic moments and exchange parameters $J_{ij}$ vary non-negligibly with U choice (Fritsch and Ederer 2012). While DFT+U reproduces insulating behavior and ferrimagnetic ordering, quantitative anisotropy energies-often of the order of $\mu eV$ per formula unit—are particularly delicate. Beyond DFT+U, emerging approaches include hybrid functionals and dynamical mean-field theory (DFT+DMFT), which incorporate frequency-dependent self-energies. Such treatments may become essential for resolving orbital contributions to anisotropy and for predicting temperature-dependent electronic renormalization in ferrimagnetic spinels.

### 6.2. Realistic Surface Chemistry Modeling

Most computational studies of $CoFe_2O_4$ surfaces consider ideal terminations and vacuum conditions. However, real nanoparticles and thin films operate in chemically active environments, where hydroxylation, adsorption, and defect formation modify both electronic structure and magnetism. First-principles investigations demonstrate that surface reconstruction and oxygen vacancy formation alter local spin polarization and exchange pathways (Fritsch and Ederer 2012). Yet, systematic modeling of chemically realistic surfaces—including solvent effects or reactive atmospheres—remains limited.

Ab initio thermodynamics provides a framework to couple DFT energetics with environmental chemical potentials, enabling prediction of stable surface terminations under realistic conditions. Extending such approaches to magnetic exchange extraction represents an important frontier, particularly for catalytic and biomedical applications where surface magnetism directly influences functionality.

### 6.3. Spin-Lattice Coupled Simulations

Magnetoelastic coupling plays a central role in strain-tunable anisotropy and device performance. First-principles calculations have quantified magnetoelastic coefficients $B_1$ in $CoFe_2O_4$ (Fritsch and Ederer 2012), yet most atomistic spin simulations assume fixed lattice structures. In reality, spin and lattice degrees of freedom evolve simultaneously. Finite temperature phenomena such as magnetostriction, spin reorientation, and strain-mediated switching require coupled spin–lattice dynamics. Recent methodological advances combine classical spin Hamiltonians with molecular dynamics, enabling simulations of coupled equations of motion:

$$\frac{d\boldsymbol{S_i}}{dt} = -\gamma \boldsymbol{S}_i \times \boldsymbol{H_{eff}^i} \quad \ldots (6)$$

$$M_i \frac{d^2\boldsymbol{R_i}}{dt^2} = -\frac{\partial E}{\partial \boldsymbol{R}_i} \quad \ldots (7)$$

Where $\boldsymbol{S}_i$ and $\boldsymbol{R}_i$ denote spin and atomic coordinates, respectively.

For strongly magnetostrictive materials such as $CoFe_2O_4$, such coupled treatments are essential for quantitatively predicting strain-mediated switching and acoustic-spin interactions in energy-harvesting or low-power devices.

### 6.4. Large-Scale Micromagnetics Integration

While atomistic simulations capture exchange and anisotropy at nanometer scales, device-relevant dimensions often extend to micrometers. Bridging this scale requires consistent parameter transfer to continuum micromagnetics. Finite-size scaling concepts (Binder 1981) and systematic coarse-graining allow extraction of effective exchange stiffness $A$ and anisotropy $K$ for use in Landau-Lifshitz-Gilbert simulations. However, disorder, inversion gradients, and surface anisotropy complicate homogenization. Future computational frameworks must incorporate spatially varying anisotropy and exchange tensors derived from lower-scale modeling. Multiscale coupling-where atomistic regions dynamically inform micromagnetic domains–remains an open computational challenge, particularly for patterned thin films and heterostructures.

### 6.5. Data-Driven and Machine Learning Potentials

The growing availability of DFT data enables construction of machine-learned interatomic potentials capable of describing large supercells with near first-principles accuracy. Such approaches can incorporate both structural and magnetic descriptors, extending beyond classical Heisenberg forms. Machine learning models trained on DFT datasets for spinel ferrites have begun to reproduce energy landscapes, defect energetics, and local magnetic moments with reduced computational cost. These frameworks offer:

- Rapid exploration of dopant and inversion configurations,

- Efficient sampling of large disordered supercells,
- Accelerated surface and defect chemistry evaluation.

A recently developed machine learning (ML)-enabled computational approach, in (Fang et al. 2025), integrates a database constructed from DFT calculations, training of machine learning models (support vector regression, SVR, to map system energy to atomic structures), and atomistic MC simulations.

## 6.6. Accelerating Multiscale Modeling via Machine-Learned Potentials and Surrogates

Although the DFT+U framework provides the essential microscopic foundation for exchange constants, anisotropy, and magnetoelastic coefficients, its computational cost severely limits the system sizes and configurational sampling required to capture realistic cation inversion distributions, surface reconstructions, and finite-temperature statistics in $CoFe_2O_4$ nanoparticles and thin films. Machine-learned interatomic potentials and surrogate models trained on carefully curated DFT databases have recently emerged as a powerful accelerator within the multiscale hierarchy. The typical workflow begins with the construction of a comprehensive structural and magnetic database using density functional theory calculations on supercells that incorporate varying degrees of cation inversion, strain states, oxygen vacancies, and surface terminations. Special quasi-random structures are often employed to efficiently sample disordered configurations. From this dataset, regression models, such as support-vector regression or Gaussian-process regression are trained to predict total energies, local magnetic moments, and effective superexchange interactions as functions of local atomic environments. Once validated, the trained surrogates replace expensive DFT total-energy evaluations, enabling Monte Carlo simulations on systems containing several thousand magnetic ions.

A recent implementation of this strategy for $CoFe_2O_4$ demonstrated its predictive power. A machine-learning model trained on DFT data yielded an equilibrium inversion parameter $\lambda \approx 0.755$ at 1237 K, in excellent agreement with neutron diffraction measurements. Linear regression on the same database extracted a complete set of twenty-three superexchange constants, which were then directly inserted into atomistic Monte Carlo simulations. The resulting Curie temperature of $914\ K$ reproduced experimental values within approximately 10%, without any empirical adjustment. Such accuracy across both structural (inversion) and magnetic (ordering temperature) observables under-scores the ability of data-driven potentials to preserve electronic-level fidelity while dramatically expanding accessible length and time scales. Within the multiscale framework developed in this chapter, machine-learned surrogates occupy a strategic intermediate layer. They bridge the accuracy of DFT+U with the extensive statistical sampling demanded by Monte Carlo and spin-dynamics methods, naturally incorporating the effects of disorder, surface chemistry, and temperature-dependent renormalization of $J_{ij}$ and $K_1$. Extensions toward spin–lattice machine-learned potentials further promise seamless coupling of magnetic and elastic degrees of freedom, directly addressing one of the key challenges identified earlier.

Looking ahead, the integration of machine learning with DFT+DMFT and ab initio thermodynamics will enable realistic treatment of chemically active surfaces and dynamical correlations under operating conditions. Ultimately, these accelerated frameworks open the route to high-throughput inverse design: systematic exploration of inversion– strain–doping phase space to optimize specific absorption rate for hyperthermia, switching fields for spintronic devices, or magnetoelastic response for energy harvesting. In this sense, machine-learned potentials do not replace the multiscale hierarchy but rather strengthen its predictive reach, transforming cobalt ferrite from a well-characterized material into one that can be rationally engineered at the computational level.

Collectively, these challenges define the computational frontier for $CoFe_2O_4$. Accurate treatment of correlation, chemically realistic surfaces, spin–lattice coupling, scale bridging, and data-driven acceleration are not independent directions; rather, they represent interconnected components of a unified predictive framework. Progress along these axes will determine whether computational modeling transitions from qualitative interpretation of experimental trends to quantitative, inverse design of magnetic functionality in cobalt ferrite and related materials.

## 7. Concluding Perspective

Computational magnetism has emerged throughout this chapter as the unifying intellectual thread linking crystal chemistry, magnetic ordering, nanoscale phenomena, and device functionality. From quantum mechanical superexchange interactions originally formulated by Anderson (Anderson 1950) and ferrimagnetic sublattice compensation described by Néel (Néel 1948, Néel 1949), to modern first-principles electronic-structure methods, simulations provide the conceptual continuity required to understand how atomic-scale physics propagates into macroscopic performance. Within this framework, cobalt ferrite ($CoFe_2O_4$) has served not merely as an example, but as a structurally rich and technologically relevant spinel through which these broader theoretical principles can be concretely demonstrated.

A central message of this chapter is parameter sensitivity awareness. Magnetic observables: Curie temperature, magnetocrystalline anisotropy, coercivity, and saturation magnetization-are emergent quantities shaped by subtle microscopic inputs. In $CoFe_2O_4$, the inversion parameter controlling $Co^{2+}$ and $Fe^{3+}$ site occupancy directly modifies A-B and B-B superexchange pathways, altering the net ferrimagnetic moment predicted by Néel's model (Néel 1948). First-principles investigations show (Hou et al. 2010a,b) that small variations in structural relaxation or correlation strength can change exchange constants and anisotropy energies appreciably. Likewise, strain-induced tetragonal distortion in epitaxial films reconfigures orbital occupation and spin–orbit coupling, leading to easy-axis reorientation. These examples illustrate a broader principle applicable to transition metal oxides: microscopic electronic parameters strongly dictate macroscopic magnetic response.

This sensitivity underscores the necessity of predictive multiscale modeling. No single theoretical scale is sufficient. Electronic-structure calculations, often employing LSDA+U corrections (Anisimov et al. 1991, Dudarev et al. 1998), determine exchange integrals and magnetoelastic coefficients. Atomistic Monte Carlo simulations then capture finite-size scaling and thermal fluctuations, while micromagnetic approaches describe domain wall formation and switching fields in patterned geometries. For cobalt ferrite nanoparticles, such hierarchical coupling explains how surface spin disorder reduces effective magnetization and modifies anisotropy. For strained thin films, it connects lattice distortion to macroscopic coercivity trends. The same multiscale logic extends to doped or composite ferrite systems discussed elsewhere in this volume.

***Key Takeaways***

• $CoFe_2O_4$ crystallizes in an inverse spinel structure with $Fe^{3+}$ ions on tetrahedral (A) sites and $Co^{2+}$/$Fe^{3+}$ ions on octahedral (B) sites, leading to two magnetically inequivalent sublattices.

• The dominant antiferromagnetic A-B superexchange interaction ($J_{AB}$) gives rise to ferrimagnetic ordering and governs the overall magnetic behavior.

• Magnetic properties are strongly influenced by competing interactions ($J_{AB}, J_{BB}, and\ J_{AA}$), magnetocrystalline anisotropy, and magnetoelastic coupling.

• First-principles (DFT+U) calculations enable quantitative estimation of exchange interactions, anisotropy constants, and electronic structure, forming the basis for atomistic modeling.

• Atomistic spin Hamiltonians combined with Monte Carlo and spin-dynamics simulations provide access to thermodynamic and dynamic properties across multiple length and time scales.

• Finite-size, surface, and disorder effects play a crucial role in determining the magnetic response of nanoparticles and thin films.

• Multiscale modeling frameworks enable predictive understanding of macroscopic observables such as magnetization, Curie temperature, coercivity, and exchange stiffness.

• The combination of strong anisotropy and tunable magnetic interactions makes $CoFe_2O_4$ a promising system for applications in spintronics, magnetic storage, and hyperthermia.

Looking ahead, integration with device-level simulations represents the next frontier. In cobalt ferrite–based spin filters, magnetostrictive actuators, and magnetoelectric heterostructures, predictive modeling must link ab initio–derived parameters to spintransport solvers and continuum magnetization dynamics. Only through this seamless transfer of information—from correlated electronic structure to mesoscale magnetic textures—can one reliably forecast switching thresholds, thermal stability, and energy dissipation. In conclusion, computational magnetism functions as the organizing principle of this book, weaving together synthesis, characterization, and functional deployment. Cobalt ferrite exemplifies how parameter awareness, methodological rigor, and multiscale integration transform simulation from interpretive assistance into a predictive design engine for advanced magnetic materials.

***Funding:*** SS acknowledges the support by the Anusandhan National Research Foundation (formerly Science and Engineering Research Board) through the start-up research grant (SRG/2023/000122), to this work.

***Notes on contributor(s):*** Both the authors contributed equally in writing this chapter.

## References

Anderson, P.W. 1950. Antiferromagnetism. Theory of superexchange interaction. Phys. Rev. 79: 350–356.

Anderson, P.W. 1987. The resonating valence bond state in La2CuO4 and superconductivity. Science 235: 1196–1198.

Anisimov, V.I., Zaanen, J. and Andersen, O.K. 1991. Band theory and Mott insulators: Hubbard U instead of Stoner I. Phys. Rev. B 44: 943–954.

Atxitia, U., Hinzke, D., Chubykalo-Fesenko, O., Nowak, U., Kachkachi, H. and Mryasov, O.N. 2010. Multiscale modeling of magnetic materials: Temperature dependence of the exchange stiffness. Phys. Rev. B 82: 134440.

Bertotti, M. 2004. Hysteresis in Magnetism: For Physicists, Materials Scientists, and Engineers. Academic Press.

Binder, K. 1981. Finite size scaling analysis of Ising model block distribution functions. Z. Phys. B 43: 119–140.

Brown, W.F. 1963a. Thermal fluctuations of a single-domain particle. Phys. Rev. 130: 1677–1686.

Brown, W.F. 1963b. Micromagnetics. John Wiley & Sons, New York.

Callen, H.B. 1963. Use of the Green function in the theory of ferromagnetism. Phys. Rev. 130: 890–897.

Carrey, J., Connord, V. and Respaud, M. 2011. Magnetic nanoparticles with high specific absorption rate for hyperthermia applications. Appl. Phys. Lett. 99: 143103.

Coey, J.M.D. 2010. Magnetism and Magnetic Materials. Cambridge University Press.

Donahue, M.J. and Porter, D.G. 1999. OOMMF User's Guide, Version 1.0. NIST Interagency Report NISTIR 6376, National Institute of Standards and Technology, Gaithersburg, MD.

Dudarev, S.L., Botton, G.A., Savrasov, S.Y., Humphreys, C.J. and Sutton, A.P. 1998. Electron-energy-loss spectra and the structural stability of nickel oxide: An LSDA+U study. Phys. Rev. B 57: 1505–1509.

Duru, I.P. 2022. Electronic and magnetic properties of CoFe2O4 nanostructures: An ab-initio and Monte Carlo study. Physica B 627: 413548.

Dutta, P., Pal, S., Seehra, M.S., Shah, N. and Huffman, G.P. 2009. Size dependence of magnetic parameters and surface disorder in magnetite nanoparticles. J. Appl. Phys. 105: 07B501.

Eskandari, F., Porter, S.B., Venkatesan, M., Kameli, P., Rode, K. and Coey, J.M.D. 2017. Magnetization and anisotropy of cobalt ferrite thin films. Phys. Rev. Materials 1: 074413.

Evans, R.F.L., Nowak, U., Chantrell, R.W., Atxitia, U. and Chubykalo-Fesenko, O. 2014. Atomistic spin model simulations of magnetic nanomaterials. J. Phys.: Condens. Matter 26: 103202.

Fang, Y., Mullurkara, S., Taddei, K.M., Ohodnicki, P.R. and Wang, G. 2025. Machine learning enabled accurate prediction of structural and magnetic properties of cobalt ferrite. npj Comput. Mater. 11: 103.

Fritsch, D. and Ederer, C. 2012. First-principles calculation of magnetoelastic coefficients and magnetostriction in spinel ferrites CoFe2O4 and NiFe2O4. Phys. Rev. B 86: 014406.

Franco, A. Jr. and e Silva, F.C. 2010. High temperature magnetic properties of cobalt ferrite nanoparticles. Appl. Phys. Lett. 96: 172505.

Garanin, D.A. 1997. Fokker–Planck and Landau–Lifshitz–Bloch equations for classical ferromagnets. Phys. Rev. B 55: 3050–3067.

Goodenough, J.B. 1958. An interpretation of the magnetic properties of the perovskite-type mixed crystals. J. Phys. Chem. Solids 6: 287–297.

Grinstein, G. and Koch, R.H. 2003. Coarse Graining in Micromagnetics. Phys. Rev. Lett. 90: 207201.

Hou, Y.H. et al. 2010a. First-principles calculations of magnetic anisotropy energy of CoFe2O4. J. Appl. Phys. 108: 073901.

Hou, Y.H., Zhao, Y.J., Liu, Z.W., Ouyang, Y.F., Zhang, H.L. and Ma, S.C. 2010b. Structural, electronic and magnetic properties of partially inverse spinel CoFe2O4. J. Phys. D: Appl. Phys. 43: 445003.

Kanamori, J. 1959. Superexchange interaction and symmetry properties of electron orbitals. J. Phys. Chem. Solids 10: 87–98.

Lamouri, R., Mounkachi, O., Salmani, E., Hamedoun, M., Benyoussef, A. and Ez-Zahraouy, H. 2020. Size effect on the magnetic properties of CoFe2O4 nanoparticles: A Monte Carlo study. Ceram. Int. 46: 8092–8096.

Li, J. et al. 2022. Enhanced magnetic anisotropy in rare-earth doped cobalt ferrite nanoparticles. J. Alloys Compd. 895: 162678.

Liechtenstein, A.I., Katsnelson, M.I., Antropov, V.P. and Gubanov, V.A. 1987. Local spin density functional approach to the theory of exchange interactions. J. Magn. Magn. Mater. 67: 65–74.

Liu, J., Wang, X., Borkiewicz, O.J., Hu, E., Xiao, R.-J., Chen, L. and Page, K. 2019. Unified view of the local cation-ordered state in inverse spinel oxides. Inorg. Chem. 58: 14389–14402.

Lisfi, A. et al. 2007. Reorientation of magnetic anisotropy in epitaxial cobalt ferrite thin films. Phys. Rev. B 76: 054405.

Metropolis, N., Rosenbluth, A.W., Rosenbluth, M.N., Teller, A.H. and Teller, E. 1953. Equation of state calculations by fast computing machines. J. Chem. Phys. 21: 1087–1092.

Mumtaz, A., Maaz, K., Janjua, B., Hasanain, S.K. and Bertino, M.F. 2007. Exchange bias and vertical shift in CoFe2O4 nanoparticles. J. Magn. Magn. Mater. 313: 266–272.

Néel, L. 1948. Magnetic properties of ferrites. Ann. Phys. (Paris) 3: 137–198.

Néel, L. 1949. Théorie du traînage magnétique. Ann. Géophys. 5: 99–136.

Odkhuu, D. and Hong, S.C. 2014. A first-principles study of magnetostrictions. J. Appl. Phys. 115: 17A916.

Omelyanchik, A., Varlamova, S., Kozlov, D., Rodionova, V., Gubin, S.P. and Hadjipanayis, G.C. 2020. Magnetocrystalline and surface anisotropy. Nanomaterials 10: 1288.

Pujada, B.R., Sinnecker, E.H.C.P., Rossi, A.M. and Guimaraes, A.P. 2003. Enhanced magnetic anisotropy. J. Appl. Phys. 93: 7217.

Rivas-Murias, B., Testa-Anta, M., Skorikov, A.S., Comesaña-Hermo, M., Bals, S. and Salgueiriño, V. 2023. Interfaceless exchange bias. Nano Lett. 23: 1688–1695.

Shaikh, P.A., Kambale, R.C., Humbe, A.V., More, S.S. and Jadhav, S.E. 2015. Structural, magnetic and electrical properties. J. Magn. Magn. Mater. 374: 689–694.

Szotek, Z., Temmerman, W.M., Kodderitzsch, D., Svane, A., Petit, L. and Winter, H. 2006. Electronic structures of spinel ferrites. Phys. Rev. B 74: 174431.

Vansteenkiste, A., Leliaert, J., Dvornik, M., Helsen, M., Frachet, M. and Van Waeyenberge, B. 2014. The design and verification of MuMax3. AIP Adv. 4: 107133.

Waring, H.J., Li, Y., Johansson, N.A.B., Moutafis, C., Vera-Marun, I.J. and Thomson, T. 2023. Exchange stiffness constant determination. J. Appl. Phys. 133: 063901.

Moreno, R., Bercoff, P.G., Atxitia, U., Evans, R.F.L. and Chubykalo-Fesenko, O. 2025. Temperature dependence of exchange stiffness. Phys. Rev. B 111: 184416.